# The security deposit for finitely repeated Prisoner's dilemma

Deng Xuegong

Northeastern university school of science, Shenyang, Liaoning, china 110004

**Abstract**

Under the assumption of complete rationality, Nash equilibrium is the only reasonable strategy (set) of the finitely repeated prisoner's dilemma. In fact, some strategies only slightly deviate from the so-called rationality, and the corresponding payoff may much better than that of Nash equilibrium. This article points out, even under the rational assumptions, the players have reason to seek a mutually beneficial agreement (Pareto dominated compare to Nash equilibrium) and a weak and optional constraints, so that the agreement can be successfully implemented. If the constraint does not harm the interests of the participants, or the adversely affects of the constraint are negligible, then the finitely repeated prisoner's dilemma becomes a bargaining problem issues on the strategy sequences and the problem to seek the constraints. The quantification of the constraints, the so-called security deposit in this paper, is nearly a concept of distance from an agreement (a strategy set) to the complete rationality.

## Background

Finitely repeated Prisoner's dilemma is theoretically attracted by the only Nash equilibrium, as long as there is not a binding agreement between the two players [1-4]. But in countless simulations, many people actually try to use the cooperative strategy even he has some knowledge of Nash equilibrium and backward reasoning [5-8]. This phenomenon cannot be entirely attributed to altruistic tendencies or moral strength [3,9,10]. Of course, the willingness to cooperate depends on the payoff matrix of the Prisoner's Dilemma and the number of repetitions. In a real finitely repeated game, sometime, both players seem to know there is a folk theorem, even though the theorem is only for infinitely repeated game[3,11]. In the process of finitely repeated Prisoner's dilemma, the folk theorem seems closer to everyman's intuition and Nash equilibrium seems a little far away from our intuition. Certainly, it is a logically unreasonable estimation.

If the players are rational, then Nash equilibrium is their only reasonable choice. By very simple backward reasoning, we can get this conclusion, even if they have privately reached an oral agreement before the game. But if all defections are due to some negligible reasons and the players can find some very weak and acceptable binding terms for their agreement, the agreement can be successfully implemented. First of all, the payoff expectation of this agreement might be better than that of the Nash equilibrium.

This paper presents a concept of a security deposit, if we do not consider the discount rate, it will not bring any damage to the participants. It indicates that sometimes an agreement with better payoff expectation than Nash equilibrium is very easy to be implemented.

**Security deposit for finitely repeated Prisoner's dilemma**

An agreement of N times repeated Prisoner's dilemma is a mutually acceptable strategy serial, $s_i^\alpha, i = 1 \cdots N$, $\alpha = 1,2$ represents player 1 and 2. In this article, it is also expressed as $s_1^1 s_2^1 \cdots s_N^1 \leftrightarrow s_1^2 s_2^2 \cdots s_N^2$, it is just to describe the correspondence of the strategy of the two players more clearly. The corresponding payoff is also denoted as $v_1^1 v_2^1 \cdots v_N^1 \leftrightarrow v_1^2 v_2^2 \cdots v_N^2$. The payoff matrix of Prisoner's Dilemma is described by Figure 1, satisfy: e> a> g> c and d> b> h> f. Following three assumptions are the bases of this article, they are, (1) Any player, at any stage, defect agreement, the agreement will enter the single-state B↔D. (2) any player will not fight for a unfavorable terms on the other player, if he is not to get benefits for his own. (3) Any effective agreement can only be Pareto dominated compare to Nash equilibrium.

|  |  | Jack | |
|---|---|---|---|
|  |  | C | D |
| Tom | A | (a, b) | (c, d) |
|  | B | (e, f) | (g, h) |

Figure 1 the schematic diagram of Prisoner's Dilemma

Because A↔C is Pareto dominated, compare to Nash equilibrium B↔D, only three kinds of strategies pairs occur in the agreement, they are, A↔C, A↔D and B↔C. Any fragment of an agreement, without considering the order of the strategies pairs, can be written as

$$A_1 \cdots A_x B_1 \cdots B_y A_1 \cdots A_z \leftrightarrow C_1 \cdots C_x C_1 \cdots C_y D_1 \cdots D_z$$

When y=0 and z=0, it is called a full cooperation fragment. When x=0, it is called a mutually beneficial fragment. When y=0 or z=0, it is called compensation fragment and when y=0 & x=0 or Z=0 & x=o, it is called strong compensation fragment. When none of x, y, z equal to zero, it is called mixed fragments.

During the implementation process of an agreement, in order to get more benefits or to avoid losses, the players might defect. So Tom and Jack agreed to pay certain amount of security deposit to a trusted third party before the start of the game. In the course of the game, if any player defect the agreement, the defector will lose his security deposit, while the game will go into a single state B↔D. The player who did not defect will get the full refund of his security deposit.

Since the security deposit is sufficiently large to prevent the defection for both players, as long as they are all rational person. Both players will get the full refund when the game ended. If we do not consider the discount rate, the players will not suffer any loss. The security deposit of any agreement or a fragment of the agreement S is denoted as $I^\alpha(S)$, $\alpha = 1\ or\ 2$ for Tom a Jack.

First, let's discuss the full cooperation fragment $A_1 \cdots A_x \leftrightarrow C_1 \cdots C_x$ which is composed by x times A↔C. For Tom, no matter which step k he choose to defect, his maximum benefit is (k-1)a+e+(m-k)g, and the minimum is (k-1)a+c+(m-k)g, if his security deposit is exactly e-c, Tom will not defect this cooperation. In Tom's standpoint, if Jack will never defect, he have no reason to defect and lost his security deposit for a lower benefit, if Jack defect at any step of the game, Tom should not defect, since he can get his refund bigger than the benefit of one time defection. So Tom will never defect, if he is a rational player. So does jack, and his security deposit is just d-f.

Then discuss the mutually beneficial fragment $A_1 \cdots A_x B_1 \cdots B_y \leftrightarrow D_1 \cdots D_x C_1 \cdots C_y$, x times A ↔D and y times B ↔C. In the first x step, Jack has no reason to defect, because there is no *benefit*. If Tom wants to defect, he should choice the first step, then his maximum profit is x(g-c), if it is Tom's security deposit, he had no reason to defect too. Similarly, Jack's security deposit is y(h-f).

Using almost the same method we can calculate the security deposit of other three type of fragments. All calculation results are listed in Table 1. It should be noted that these security deposit is only to ensure that the two players in these fragments will not defect the game, do not ensure that the entire course of the game will not be defected.

| Types | Fragments of agreement | Tom's security deposit | Jack's security deposit |
|---|---|---|---|
| 1 | $A_1 \cdots A_x \leftrightarrow C_1 \cdots C_x$ | $e - c$ | $d - f$ |
| 2 | $A_1 \cdots A_x B_1 \cdots B_y \leftrightarrow D_1 \cdots D_x C_1 \cdots C_y$ | $x(g - c)$ | $y(h - f)$ |
| 3 | $A_1 \cdots A_x B_1 \cdots B_y \leftrightarrow C_1 \cdots C_{x+y}$ | $e - c$ | $(d - f) + y(h - f)$ |
| 4 | $A_1 \cdots A_{x+z} \leftrightarrow C_1 \cdots C_x D_1 \cdots D_z$ | $(e - c) + z(g - c)$ | $d - f$ |

Table 2，the security deposit of 4 types of fragments

Proposition 1: Let $S$ be an effective agreement, and S is divided into some fragments as $S = S_1 S_2 \cdots S_n$, each $S_k$ is Pareto dominant compare to the Nash equilibrium. The security deposits of two players are denoted as $I^\alpha(S_k)$ then

$$I^\alpha(S) = \underset{k \leq n}{Max}(I^\alpha(S_k))$$

is a security deposit of the full agreement S.

Proof: We can use the backward reasoning to prove this proposition.

Let $v_i^\alpha$ be the payoff expectation of $S_i$. $r_{i,(*,*)}^\alpha$ be the real payoff, (*,*) denote the situation of Tome and jack, 0 means corporation and 1 means defection. $e_i^\alpha$ be payoff of Nash equilibrium of the fragment. If the game is maintained until the final stage $S_n$. if Tom ($\alpha = 1$) defect, his payoffs is

$$\sum_{i=1}^{n-1} v_i^1 + r_{n,(1,*)}^1 - I^1(S)$$

That is to say he will lost his full security deposit, but he knows his maximum benefit of defection in $S_n$ will not be greater than $I^\alpha(S_n)$, not to mention $I^1(S)$. So Tom will not defect in the final stage. The same is true for Jack.

In the stage $S_{n-1}$, if Tom defects, his total payoff is

$$\sum_{i=1}^{n-2} v_i^1 + r_{n-1,(1.*)}^1 - I^1(S) + e_n^1$$

If tom does not defect, but Jack defect, his total payoff is

$$\sum_{i=1}^{n-2} v_i^1 + r_{n-1,(0.1)}^1 + e_n^1$$

If both Tom and Jack do not defect, since they will not defect in next step, Tom's total payoff is

$$\sum_{i=1}^{n-2} v_i^1 + r_{n-1,(0.0)}^1 + v_n^1$$

As the definition of $I^1(S)$, there are two relations

$$r_{n-1,(0.1)}^1 \geq r_{n-1,(1.1)}^1 - I^1(S)$$
$$r_{n-1,(0.0)}^1 \geq r_{n-1,(1.0)}^1 - I^1(S)$$

There is also $v_n^1 \geq e_n^1$. So Tom will not defect in the stage n-1, so in the whole game. The same is true for Jack. On the other word, $I^1(S)$ is a security deposit of the full agreement S.

Proposition 2: Let $S$ be an effective agreement, and S is divided into some fragments as $S = S_1 S_2 \cdots S_n$, in which, $S_1$ the only is *full cooperation fragment*, $S_2$ to $S_{n-1}$ are *mutually beneficial fragments*, each them is Pareto dominant compare to the Nash equilibrium. Only $S_n$ is not Pareto dominant compare to the Nash equilibrium (but can be empty), then

$$I^\alpha(S) = Max_{i<n}(I^\alpha(S_i)) + I^\alpha(S_n)$$

is a security deposit of the full agreement S.

Proposition 3: The security deposit of an agreement $S = S_1 S_2 \cdots S_n$ could be gradually refunded. If none of the players defected in the first k stages, then the security deposit dropped to

$$I_k^\alpha(S) = Max_{i \geq k}(I^\alpha(S_i)) + I^\alpha(S_n).$$

### Examples

Figure 2 is a valuation of the payoff matrix of the Prisoner's Dilemma in Figure 1, the repetition number is 29. The total number of possible agreement is 84, the number of effective agreement is 49, which are Pareto dominant compare to Nash equilibrium. Among the 49 effective agreements, there are 30 agreements with better payoffs of two player's expectation than 7.5, and 12 better than 8 and 1 better than 8.5. In any effective agreement, A↔C does not appear more than 3 times, because two B↔C plus one A↔D is better than 3 A↔C.    Partial agreements are listed and analyzed in table 2.

| Orde | Number | | | Tom | | | Jack | | |
|---|---|---|---|---|---|---|---|---|---|
| | B↔C | A↔D | A↔C | E(payoff) | Total | SD | E(payoff) | Total | SD |
| 1 | 10 | 19 | 0 | 6.069 | 176 | 4 | 17.103 | 496 | 4 |
| 2 | 11 | 17 | 2 | 6.414 | 178 | 6 | 15.862 | 452 | 20 |
| 3 | 12 | 15 | 2 | 6.759 | 180 | 6 | 14.621 | 408 | 20 |
| 4 | 15 | 14 | 0 | 7.103 | 206 | 2 | 13.655 | 396 | 4 |
| 5 | 16 | 12 | 1 | 7.448 | 208 | 6 | 12.414 | 352 | 20 |
| 6 | 17 | 10 | 2 | 7.793 | 210 | 6 | 11.172 | 308 | 20 |
| 7 | 20 | 9 | 0 | 8.138 | 236 | 2 | 10.207 | 296 | 6 |
| 8 | 22 | 7 | 0 | 8.552 | 248 | 2 | 8.828 | 256 | 8 |
| 9 | 23 | 4 | 2 | 9.034 | 246 | 6 | 7.034 | 188 | 20 |
| 10 | 26 | 3 | 0 | 9.379 | 272 | 2 | 6.069 | 176 | 18 |

Table 2. The list of 10 effective agreements, and the corresponding payoff expectations and security deposits. E(payoff)=payoff expectations of one time repetition, Total payoff= Total payoff of the whole game, SD= Security deposit

Take the 5th in Table 1 for example; it can be divided into following fragments, one A↔C, four BBA↔CCD, and eight BA↔CD. It is easy to verify that each fragment are *Pareto dominant compare to Nash equilibrium*. So in the second step of the game, the security deposit dropped from 6 and 20 to 2 and 4, after the 13th step it dropped to 2 and 2.

| | | Jack | |
|---|---|---|---|
| | | C | D |
| Tom | A | (8,8) | (4,24) |
| | B | (10,4) | (6,6) |

Figure 2, a valuation of Fig 1.

### Discussion

An effective agreement is actually a result of bargaining, the bargaining contains two meanings, one of which is to find mutual acceptable agreement (payoff), and the second is to find a relatively weak and sufficient condition to maintain the agreement. Of course, the security deposit is also an important factor that may also affect the player's behavior in the bargaining.

The security deposit in this article may not be optimal and there may be a better way to calculate it. This is a problem to be solved. Another problem is whether Prisoner's dilemma is the only original game suite to the method of this paper.

Security deposit is not intended to guarantee the players to get the maximum benefit, it only to ensure that players do not deviate from agreement. In the previous section it gives a method to find feasible algorithms for Security deposit, but they are not certainly the best results. The same as the bargaining of payoff, the Security deposit in the agreements may also be a bargaining issue. In this paper, the finitely repeated Prisoner's dilemma game comes down to bargaining issues, a more intractable problem.